\documentclass[12pt]{iopart}

\usepackage{graphicx,epsfig,floatflt}
\usepackage{amssymb}
\usepackage{iopams}
\usepackage{float}
\usepackage{multirow}

\def\beq{\begin{equation}}
\def\eeq{\end{equation}}
\def\br{\begin{eqnarray}}
\def\er{\end{eqnarray}}
\def\benu{\begin{enumerate}}
\def\eenu{\end{enumerate}}

\def\l{\left}
\def\r{\right}
\def\f{\frac}

\begin{document}

%%%%%%%%%%%%%%%%%%%%%%%%%%%%%%%%%%%%%%%%%%%%%%%%%%%%%%%%%%%%%%%%%%%%%%%%%%%%%%%
\title{Primordial features due to a step in the inflaton potential}
\author{Dhiraj Kumar Hazra$^{1+}$, Moumita Aich$^{2\dag}$, 
Rajeev Kumar Jain$^{3\ast}$,\\
L.~Sriramkumar$^{1\ddag}$ and Tarun Souradeep$^{2\S}$}
%%%%%%%%%%%%%%%%%%%%%%%%%%%%%%%%%%%%%%%%%%%%%%%%%%%%%%%%%%%%%%%%%%%%%%%%%%%%%%%
\address{$^{1}$Harish-Chandra Research Institute, Chhatnag Road,
Jhunsi,\\
${}$~Allahabad~211~019, India.}
\address{$^{2}$IUCAA, Post Bag 4, Ganeshkhind, Pune 411 007, India.}
\address{$^{3}$Department of Theoretical Physics, University of Geneva,
24~Quai Ernest-Ansermet,\\ 
${}$~CH-1211, Geneva 4, Switzerland.}
\eads{\mailto{$^{+}$dhiraj@hri.res.in}, 
\mailto{$^{\dag}$moumita@iucaa.ernet.in},\\ 
{\hskip 13mm}\mailto{$^{\ast}$rajeev.jain@unige.ch},
\mailto{$^{\ddag}$sriram@hri.res.in},\mailto{$^{\S}$tarun@iucaa.ernet.in}}
%%%%%%%%%%%%%%%%%%%%%%%%%%%%%%%%%%%%%%%%%%%%%%%%%%%%%%%%%%%%%%%%%%%%%%%%%%%%%%%
\begin{abstract}
Certain oscillatory features in the primordial scalar power spectrum are 
known to provide a better fit to the outliers in the cosmic microwave 
background data near the multipole moments of $\ell=22$ and $40$.
These features are usually generated by introducing a step in the popular,
quadratic potential describing the canonical scalar field. 
Such a model will be ruled out, if the tensors remain undetected at a level 
corresponding to a tensor-to-scalar ratio of, say, $r\simeq 0.1$.
In this work, in addition to the popular quadratic potential, we investigate 
the effects of the step in a small field model and a tachyon model.
With possible applications to future datasets (such as PLANCK) in mind, we 
evaluate the tensor power spectrum exactly, and include its contribution in 
our analysis. 
We compare the models with the WMAP (five as well as seven-year), the QUaD 
and the ACBAR data. 
As expected, a step at a particular location and of a suitable magnitude and 
width is found to improve the fit to the outliers (near $\ell=22$ and $40$) 
in all these cases.
We point out that, if the tensors prove to be small (say, $r\lesssim 0.01$), 
the quadratic potential and the tachyon model will cease to be viable, and 
more attention will need to be paid to examples such as the small field models.
\end{abstract}
\maketitle

%%%%%%%%%%%%%%%%%%%%%%%%%%%%%%%%%%%%%%%%%%%%%%%%%%%%%%%%%%%%%%%%%%%%%%%%%%%%%%%

\section{Outliers, inflationary models with a step, and tensors}

Inflation continues to remain the most promising paradigm for describing 
the origin of the perturbations in the early universe.
It has been performing remarkably well against the observational data, and
the challenge before the other competing scenarios is to match the simplicity 
and efficiency of inflation.
Many models of inflation lead to an epoch of slow roll lasting for, say, 
$50$-$60$ e-folds, as is required to resolve the horizon problem.
It is well known that slow roll inflation leads to a featureless, nearly 
scale invariant, power law, primordial scalar spectrum.
Such a spectrum, along with the assumption of a spatially flat, concordant 
$\Lambda$CDM background cosmological model, provides a good fit to the 
recent observations of the anisotropies in the Cosmic Microwave Background 
(CMB) by different missions such as the Wilkinson Microwave Anisotropy Probe 
(WMAP)~\cite{wmap-5,wmap-7}, the QUEST at DASI (QUaD)~\cite{quad-2009}, and
the Arcminute Cosmology Bolometer Array Receiver (ACBAR)~\cite{acbar-2008}.
 
The efficacy of the inflationary scenario also seems to be responsible 
for an important drawback.
Though, as a paradigm, inflation can be considered to be a success, it 
would be fair to say that we are rather far from converging on a specific 
model or even a class of models of inflation.  
There exist a plethora of inflationary models that remain consistent with 
the data.
We mentioned above that a nearly scale invariant, power law, scalar 
spectrum fits the observations of the anisotropies in the CMB quite well.
However, there exist a few data points at the lower multipoles---notably, 
at the quadrapole ($\ell=2$) and near the multipole moments of $\ell=22$ 
and $40$---which lie outside the cosmic variance associated with the power 
law primordial spectrum.
Needless to add, statistically, a few outliers in a thousand or so data 
points can always be expected.
These outliers were noticed in the WMAP first-year data, and they continue 
to be present even in the most recent, seven-year data, making them unlikely 
to be artifacts of data analysis. 
It is possible that they actually indicate certain non-trivial inflationary 
dynamics.
In that case, these outliers are important from the phenomenological 
perspective of attempting to constrain the models from the data, because 
only a more restricted class of inflationary models can be expected to 
provide an improved fit to these outliers.
Therefore, it is a worthwhile exercise to systematically explore models that 
lead to specific deviations from the standard power law, inflationary 
perturbation spectrum, and also provide an improved fit to the data.

Various efforts towards a model independent reconstruction of the primordial 
spectrum from the observed pattern of the CMB anisotropies seem to indicate 
the presence of certain features in the spectrum~\cite{rc}.
(However, we should add that there also exist other views on the 
possibility of features in the primordial spectrum; in this context, 
see, for example, Refs.~\cite{rc-wf}.)
In particular, a burst of oscillations in the primordial spectrum seems 
to provide a better fit to the CMB angular power spectrum near the 
multipole moments of $\ell=22$ and $40$.
Generating these oscillations requires a short period of deviation from 
slow roll inflation~\cite{starobinsky-1992,dvorkin-2010}, and such a 
departure has often been achieved by introducing a small step in the 
popular, quadratic potential describing the canonical scalar field 
(see Refs.~\cite{adams-2001,wmap-1,covi-2006-2007,mortonson-2009}; 
for a discussion 
on other models, see, for instance, Refs.~\cite{joy-2008-2009,jain-2009}).
At the cost of three additional parameters which characterize the location,
the height and the width of the step, it has been found that this model 
provides a considerably better fit to the CMB data with the least squares 
parameter $\chi_{\rm eff}^{2}$ typically improving by about $7$, 
when compared to the nearly scale invariant spectrum that would have 
resulted in the absence of the step~\cite{covi-2006-2007,mortonson-2009}.
But, such a chaotic inflation model leads to a reasonable amount of tensors,
and these models will be ruled out if tensors are not detected corresponding 
to a tensor-to-scalar ratio of, say, $r\simeq 0.1$.

Our aims in this paper are twofold.
Firstly, we wish to examine whether, with the introduction of a 
step, other inflationary models too perform equally well against 
the CMB data, as the quadratic potential does.
Secondly, we would also like to consider a model that leads to a 
tensor-to-scalar ratio of $r<0.1$, so that suitable alternative 
models exist if the tensor contribution turns out to be smaller.
Motivated by these considerations, apart from revisiting the popular 
quadratic potential, we shall investigate the effects of the step in 
a small field model~(in this context, see, for example, 
Ref.~\cite{efstathiou-2006}) and a tachyon model~\cite{steer-2004}.
Also, with possible applications to future datasets in mind (such as 
the ongoing PLANCK mission~\cite{planck}), we shall evaluate the 
tensor power spectrum exactly, and include its contribution in our 
analysis. 
We shall compare the models with the CMB data from the WMAP, the QUaD
and the ACBAR missions.
We shall consider the five as well as the seven-year WMAP 
data~\cite{wmap-5,wmap-7}, the QUaD June 2009 data~\cite{quad-2009} and
the ACBAR 2008 data~\cite{acbar-2008} to arrive at the observational
constraints on the inflationary parameters.
We find that, as one may expect, a step at a suitable location and
of a certain magnitude and width improves the fit to the outliers 
(near $\ell=22$ and $40$) in all the cases.
We point out that, if the amplitude of the tensors prove to be small, the 
quadratic potential and the tachyon model will become inviable, and we will 
have to turn our attention to examples such as the small field models.

The remainder of this paper is organized as follows.
In the following section, we shall outline the different inflationary 
models that we shall be considering in this work.
In Section~\ref{sec:method}, we shall discuss the methodology that we 
adopt for comparing the inflationary models with the data, the datasets 
that we use for our analysis, and the priors on the various parameters 
that we work with.
In Section~\ref{sec:results}, we shall present the results of our 
comparison of the theoretical CMB angular power spectra that arise 
from the various models with the WMAP five-year as well as seven-year 
data, the QUaD and the ACBAR data.
We shall tabulate the best fit values that we obtain on the background 
cosmological parameters and the parameters describing the inflationary 
models. 
We shall also illustrate the constraints that we arrive at on the 
parameters describing the step in the case of the small field model. 
Further, we shall explicitly show that the models with the step 
perform better against the data because of the fact that they 
lead to an improvement in the fit to the outliers around $\ell=22$ 
and $40$.
In Section~\ref{sec:sps}, we shall illustrate the scalar power spectra 
and the CMB angular power spectra corresponding to the best fit values
of the parameters of some of the models that we consider. 
Finally, in Section~\ref{sec:summary}, we shall close with a brief summary, 
and a few comments on certain implications of our results.

Note that we shall work in units such that $\hbar=c=(8\,\pi\,G)=1$.
Moreover, we shall assume the background cosmological model to be the 
standard, spatially flat, $\Lambda$CDM model.

%%%%%%%%%%%%%%%%%%%%%%%%%%%%%%%%%%%%%%%%%%%%%%%%%%%%%%%%%%%%%%%%%%%%%%%%%%%%%%%

\section{The inflationary models of our interest}

In this section, we shall list the different inflationary models that 
we shall consider, and briefly outline the parameters involved in each
of these cases. 

%%%%%%%%%%%%%%%%%%%%%%%%%%%%%%%%%%%%%%%%%%%%%%%%%%%%%%%%%%%%%%%%%%%%%%%%%%%%%%%

\subsection{The conventional, power law case}

When the tensor perturbations are also taken into account, the power law, 
scalar and tensor spectra are written as (see, for example, 
Refs.~\cite{bassett-2006,sriram-2009})
\beq
{\cal P}_{_{\rm S}}(k)
=A_{_{\rm S}} \l(\f{k}{k_{0}}\r)^{n_{_{\rm S}}-1}
\quad{\rm and}\quad
{\cal P}_{_{\rm T}}(k)
=A_{_{\rm T}} \l(\f{k}{k_{0}}\r)^{n_{_{\rm T}}},
\label{eq:ps-pl}
\eeq
respectively.
The quantities $A_{_{\rm S}}$ and $A_{_{\rm T}}$ denote the amplitude 
of the scalar and tensor spectra, while $n_{_{\rm S}}$ and $n_{_{\rm T}}$
denote the corresponding spectral indices.
The quantity $k_{0}$ is the pivot scale at which the amplitudes of the 
power spectra are quoted.
It is this power law case that will act as our reference model with 
respect to which we shall compare the performance of the other models 
against the data.

Given the scalar and tensor spectra, i.e. 
${\cal P}_{_{\rm S}}(k)$ and ${\cal P}_{_{\rm T}}(k)$, the tensor-to-scalar
ratio is defined as
\beq
r(k)=\f{{\cal P}_{_{\rm T}}(k)}{{\cal P}_{_{\rm S}}(k)}.
\eeq
When comparing the power law case with the observations, it is the 
tensor-to-scalar ratio~$r$ that is usually considered instead of the 
tensor amplitude $A_{_{\rm T}}$.
Often, the following slow roll consistency condition is further assumed
(see, for instance, Refs.~\cite{wmap-5,wmap-7,quad-2009,acbar-2008})
\beq
r=-8\; n_{_{\rm T}},\label{eq:cc}
\eeq
so that the power law case is basically described by the three parameters 
$A_{_{\rm S}}$, $n_{_{\rm S}}$ and~$r$.
We should mention here that we shall {\it not}\/ impose the above consistency 
condition while comparing the power law case with the data, and we shall work 
with all the four parameters $A_{_{\rm S}}$, $n_{_{\rm S}}$, $r$ and 
$n_{_{\rm T}}$.

%%%%%%%%%%%%%%%%%%%%%%%%%%%%%%%%%%%%%%%%%%%%%%%%%%%%%%%%%%%%%%%%%%%%%%%%%%%%%%%

\subsection{Canonical scalar field models}

We shall work with two types of canonical scalar field models.
We shall firstly revisit the large field, quadratic model described 
by the potential 
\beq
V(\phi) = \frac{1}{2}\,m^2\,\phi^2,\label{eq:qp}
\eeq
where $m$ represents the mass of the inflaton.
The parameter $m$ is essentially determined by the amplitude of the
scalar power spectrum. 
To achieve the required number, say, $60$ e-folds of inflation, in 
such a model, the field has to start at a suitably large value (in 
units of the Planck mass).
The field rolls down the potential, and inflation ends as the 
field nears the minimum of the potential (see, for instance, 
Refs.~\cite{bassett-2006,sriram-2009}).

The small field inflationary models offer an important alternative to 
the large field models.
In fact, in certain cases, the small field models are possibly better 
motivated from the high energy physics perspective than the simple 
large field models  (see, for example, Refs.~\cite{kinney-1995-96}).
Therefore, in addition to the quadratic potential mentioned above, 
we shall consider the small field model governed by the potential
\beq
V(\phi) = V_{0}\, \l[1-\l(\phi/\mu\r)^{p}\r].
\label{eq:sfm}
\eeq
The field starts at small values in such models, and inflation is 
terminated naturally as the field approaches the value~$\mu$. 
As is well known (and, as we shall also discuss), the quadratic
potential Eq.~(\ref{eq:qp}) leads to a tensor-to-scalar ratio of 
$r\simeq 0.1$~\cite{mortonson-2009}.
The small field model Eq.~(\ref{eq:sfm}) can lead to a smaller 
tensor-to-scalar ratio for suitable values of $\mu$ and $p$ 
(in this context, see Ref.~\cite{efstathiou-2006}).
Also, when $p<4$, the model is known to result in a substantial red 
tilt. 
We find that, if we choose $p=4$ and $\mu=15$, the model leads to a 
tilt that is consistent with observations, and a tensor-to-scalar 
ratio of $r\simeq 0.01$.
So while comparing with the data, we shall work with these specific
values of $p$ and~$\mu$, and vary~$V_{0}$.

%%%%%%%%%%%%%%%%%%%%%%%%%%%%%%%%%%%%%%%%%%%%%%%%%%%%%%%%%%%%%%%%%%%%%%%%%%%%%%%

\subsection{Tachyon model}

Tachyonic inflationary potentials are usually written in terms of 
two parameters, say, $\lambda$ and $\phi_{\ast}$, in the following 
form~\cite{steer-2004}:
\beq
V(\phi)=\lambda\; V_{1}(\phi/\phi_{\ast}),
\eeq
where $V_{1}(\phi/\phi_{\ast})$ is a function which has a maximum at 
the origin and vanishes as $\phi\to\infty$.
The tachyon model that we shall consider is described by the potential
\beq
V(\phi)=\frac{\lambda}{{\rm cosh}\, (\phi/\phi_{\ast})}.
\label{eq:tm}
\eeq
In such a potential, inflation typically occurs around $\phi\simeq 
\phi_{\ast}$.
The field rolls down the potential, and inflation ends at suitably 
large values of the field.
It is found that the quantity $(\lambda\, \phi_{\ast}^2)$ has to be 
much larger than unity to ensure that inflation lasts for a sufficiently 
long time~\cite{steer-2004}.
We find that the amplitude of the scalar perturbations is more sensitive
to $\phi_{\ast}$ than $\lambda$.
Hence, while comparing with the data, we fix the value of~$\lambda$, 
and vary $\phi_{\ast}$.
We shall set $\lambda=8.9\times 10^{-4}$.
We shall then choose the priors on $\phi_{\ast}$ such that $(\lambda\, 
\phi_{\ast}^2)$ is relatively large in order to achieve the required 
duration of slow roll inflation.

%%%%%%%%%%%%%%%%%%%%%%%%%%%%%%%%%%%%%%%%%%%%%%%%%%%%%%%%%%%%%%%%%%%%%%%%%%%%%%%

\subsection{Introduction of the step}

Given a potential, say, $V(\phi)$, we shall introduce the step by 
{\it multiplying}\/ the potential by the following function:
\beq
V_{\rm step}(\phi)
= \l[1 + \alpha\, \tanh\l(\frac{\phi-\phi_{0}}{\Delta\phi}\r)\r],
\label{eq:step}
\eeq
as is often done in the 
literature~\cite{adams-2001,wmap-1,covi-2006-2007,mortonson-2009}.
It should be pointed out that the quantity $\alpha$ is positive in the
case of the quadratic potential Eq.~(\ref{eq:qp}), whereas it is negative 
in the cases of the small field model Eq.~(\ref{eq:sfm}) and the tachyon
model Eq.~(\ref{eq:tm}).
Evidently, $\alpha$ denotes the height of the step, $\phi_{0}$ its location, 
and $\Delta\phi$ its width.
When comparing with the data, in addition to the potential parameters, we
shall vary these three parameters, along with the background cosmological 
parameters, to arrive at the observational constraints.

%%%%%%%%%%%%%%%%%%%%%%%%%%%%%%%%%%%%%%%%%%%%%%%%%%%%%%%%%%%%%%%%%%%%%%%%%%%%%%%

\section{Methodology, datasets, and priors}\label{sec:method}

We evaluate the inflationary power spectra numerically. 
In addition to the scalar power spectrum, we evaluate the tensor 
spectrum exactly, and include it in our analysis.  
We evaluate the spectra using a fast and accurate FORTRAN~90 code, which 
is divided into two parts.
The first part uses the fourth order Runge-Kutta algorithm~\cite{nr} to 
solve the background equations for the scale factor and the scalar field,
using e-folds as the independent variable. 
The initial values of the field and its velocity (i.e. $\phi$ 
and~${\dot \phi}$) are chosen by hand, and we ensure that the field 
starts in a slow roll phase.
As is the standard practice, we choose the initial value of the scale 
factor such that the pivot scale (viz. $k=0.05\; {\rm Mpc}^{-1}$) 
leaves the Hubble radius at $50$~e-folds before the end of the 
inflation~\cite{covi-2006-2007,mortonson-2009}.	
The second part of the code uses the Bulirsch-Stoer algorithm~\cite{nr} 
to evolve the perturbations.
We impose the standard Bunch-Davies initial conditions on the perturbations.
As is usually done~\cite{adams-2001,salopek-1989,ringeval-2008}, the 
initial conditions on the modes are imposed when they are well inside 
the Hubble radius~$H^{-1}$ [say, when $k=(100\; a\,H)$].
The modes are evolved until they are sufficiently outside the Hubble radius 
and the curvature perturbation reaches its asymptotic value [typically, this 
occurs when $k\simeq \l(10^{-5}\; a\,H\r)$]. 
The tensor perturbations are evolved in a similar fashion.
We should mention here that, while the speed of propagation of 
the curvature perturbations induced by the canonical scalar field
is a constant (and, equal to unity), it changes with time in the 
case of the tachyon models~\cite{steer-2004,jain-2007}.
We have carefully taken this point into account, while imposing the 
initial conditions on the modes as well as when evolving them from the 
sub-Hubble to the super-Hubble scales.

In order to arrive at the constraints on the various background 
cosmological parameters and the parameters describing the inflaton
potential, we perform a Markov Chain Monte-Carlo sampling of the 
parameter space. 
To do so, we make use of the publicly available CosmoMC 
package~\cite{cosmomc,lewis-2002}, which in turn uses the CMB anisotropy 
code CAMB~\cite{lewis-2000,camb} to generate the CMB angular power 
spectra from the primordial scalar and tensor spectra.
We evaluate the scalar and the tensor spectra for all the modes that 
are required by CAMB to arrive at the CMB angular power spectra. 
For our analysis, we consider the following CMB datasets:~the WMAP 
five-year~\cite{wmap-5} and seven-year data~\cite{wmap-7}, the QUaD 
June~2009 data~\cite{quad-2009}, and the the ACBAR~2008 
data~\cite{acbar-2008}. 
We have separately compared the models with the WMAP five-year and 
seven-year data.
We have also compared the models with the WMAP five-year data along 
with the QUaD data, and with the QUaD as well as the ACBAR data.
We have used the October~2009 version of CosmoMC (and CAMB) while 
comparing with the WMAP five-year and the QUaD and the ACBAR datasets.
When comparing with the WMAP seven-year data, we have made use of the 
more recent version (i.e. the January 2010 version) of CosmoMC and CAMB.
 
In our analysis, we take gravitational lensing into account.
Note that, to generate highly accurate lensed CMB spectra, CAMB
requires $\ell_{\rm max~scalar} \simeq (\ell_{\rm max} + 500)$, 
where $\ell_{\rm max}$ is, say, the largest multipole moment for 
which the data is available.
The WMAP data is available up to  $\ell \simeq 1200$, the QUaD data
goes up to $\ell \simeq 2000$, while the ACBAR data is available 
up to $\ell \simeq 2700$.
So, we set $\ell_{\rm max~scalar}=2500$ when dealing with the WMAP 
and the QUaD datasets and, when we include the ACBAR data, we set 
$\ell_{\rm max~scalar}=3300$.
Since the datasets involve rather large multipole moments (say, 
$\ell\gtrsim 1000$), we also take into account the Sunyaev-Zeldovich 
effect, and marginalize over the $A_{_{\rm SZ}}$ parameter.
For the power law case, we set the pivot scale to be $k_{0} = 0.05\; 
{\rm Mpc}^{-1}$ [cf. Eq.~(\ref{eq:ps-pl})].
We have made use of the publicly available WMAP likelihood code from 
the LAMBDA web-site~\cite{lambda} to determine the performance of the
models against the data.
We have have set the Gelman and Rubin parameter $\vert R-1\vert$ to be 
$0.03$ for convergence.
Lastly, we should add that we have used the Gibbs option (for the CMB $TT$ 
spectrum at the low multipoles) in the WMAP likelihood code to evaluate 
the least square parameter~$\chi_{\rm eff}^{2}$. 

As we had mentioned earlier, we incorporate the tensor perturbations 
in our analysis. 
Recall that, in the power law case, when the consistency 
condition Eq. (\ref{eq:cc}) is not imposed, the tensor power spectrum is 
described in terms of the tensor-to-scalar ratio $r$ and the tensor 
spectral index~$n_{_{\rm T}}$. 
They need to be specified along with the scalar amplitude $A_{_{\rm S}}$ 
and the corresponding spectral index $n_{_{\rm S}}$, in order to 
completely describe the primordial spectra.
We should emphasize that, in the other inflationary models, once the 
parameters that govern the potential have been specified, no further
parameters are required to describe the tensor power spectrum. 
The potential parameters determine the amplitude and shape of {\it both}\/ 
the scalar and the tensor spectra.

As we mentioned, we assume the background to be a spatially flat, 
$\Lambda$CDM model described by the four standard parameters, viz. 
$(\Omega_{\rm b}\, h^2)$ and $(\Omega_{\rm c}\, h^2)$, which represent
the baryon and CDM densities (with $h$ being related to the Hubble 
parameter), respectively, the ratio of the sound horizon to the angular 
diameter distance at decoupling $\theta$, and $\tau$ which denotes the 
optical depth to reionization.
We work with the following priors on these parameters:
$0.005 \le (\Omega_{\rm b}\, h^2) \le 0.1$,  $0.01 \le (\Omega_{\rm c}\, 
h^2) \le 0.99$, $0.5 \le \theta \le 10.0$ and $0.01 \le \tau \le 0.8$.
We should add that we keep the same priors on the background parameters
for all the models and datasets that we consider in our analysis.
In Table~\ref{tab:priors-ip}, we have listed the priors that we choose 
on the different parameters which describe the various inflationary models 
that we consider. 
%%%%%%%%%%%%%%%%%%%%%%%%%%%%%%%%%%%%%%%%%%%%%%%%%%%%%%%%%%%%%%%%%%%%%%%%%%%%%%%
\begin{table}[!htb]
\begin{scriptsize}
\begin{center}
\begin{tabular}{|c|c|c|c|}
\hline\hline\hline
Models & Parameter & Lower limit & Upper limit \\
\hline\hline\hline
& ${\rm ln}\, \l[10^{10}\, A_{_{\rm S}}\r]$  & $2.7$   & $4.0$\\
\cline{2-4}
Power law&$n_{_{\rm S}}$ & $0.5$ & $1.5$\\
\cline{2-4}
case & $r$ & $0.0$ & $1.0$\\
\cline{2-4}
& $n_{_{\rm T}}$ & $-0.5$ & $0.5$\\
\hline\hline
& ${\rm ln}\, \l[10^{10}\, m^2\r]$  & $-0.77$   & $-0.58$\\
\cline{2-4}
Quadratic model & $\alpha$ & $1.3\times10^{-3}$ & $1.7\times10^{-3}$\\
\cline{2-4}
with a step & $\phi_0$ & $13.0$ & $15.0$\\
\cline{2-4}
&$\Delta \phi$ & $0.015$ & $0.03$\\
\hline\hline
& ${\rm ln}\,  \l[10^{10}\,V_{0}\r]$ &$1.50$&$1.86$   \\
\cline{2-4}
Small field model & $-\alpha$ &$1.0 \times 10^{-4}$& $2.0\times10^{-4}$\\
\cline{2-4}
with a step & $\phi_0$ &$7.8$ & $8.1$ \\
\cline{2-4}
&$\Delta \phi$ &$5.0 \times 10^{-3}$ & $1.0 \times 10^{-2}$\\
\hline\hline
& ${\rm ln}\, \l[10^{10}\, \phi_{\ast}\r]$  & $34.506$   & $34.518$\\
\cline{2-4}
Tachyon model & $-\alpha$ & $1.3 \times 10^{-3}$   & $1.9\times 10^{-3}$\\
\cline{2-4}
with a step & $\phi_{0}$ & $7.81 \times 10^{5}$   & $7.83 \times 10^{5}$\\
\cline{2-4}
&$\Delta \phi$ & $340$   & $410$\\
\hline\hline\hline
\end{tabular}
\caption{\label{tab:priors-ip}The priors on the various parameters 
that describe the primordial spectrum in the power law case, and the 
inflationary potential in all the other cases.
We work with these priors when comparing the models with all the 
datasets.}
\end{center}
\end{scriptsize}
\end{table}
%%%%%%%%%%%%%%%%%%%%%%%%%%%%%%%%%%%%%%%%%%%%%%%%%%%%%%%%%%%%%%%%%%%%%%%%%%%%%%%

\section{Bounds on the background cosmology and the inflationary 
parameters}\label{sec:results}

In this section, we shall present the results of our analysis.
As we mentioned above, we have separately compared the models with 
the WMAP five-year and seven-year data.
We have also compared the models with the WMAP five-year data along 
with the QUaD data, and with the QUaD as well as the ACBAR data.
We shall tabulate below the best fit values that we arrive at on the 
various background and inflationary parameters for the power law case 
and for the inflationary models with the step.
We shall also illustrate the behavior of the one-dimensional likelihoods
for the three parameters (i.e. $\alpha$, $\phi_{0}$ and $\Delta\phi$)
that describe the step in the case of the small field model Eq. (\ref{eq:sfm}).
Further, we shall also provide the least squares parameter 
$\chi_{\rm eff}^{2}$ in all the cases, viz. the power law case and the
three inflationary models, with and without the step.

%%%%%%%%%%%%%%%%%%%%%%%%%%%%%%%%%%%%%%%%%%%%%%%%%%%%%%%%%%%%%%%%%%%%%%%%%%%%%%%

\subsection{The best fit values and the one-dimensional likelihoods}

In Tables~\ref{tab:pl} and \ref{tab:am-ws}, we have listed the best fit 
values for the various parameters in the power law case and in the three
inflationary models with the step.
These tables contain the results that we arrive at upon comparing the 
models with the WMAP five-year (denoted as WMAP-$5$) and seven-year
(denoted as WMAP-$7$) data, the QUaD as well as the ACBAR data sets.
Note that we have only presented the results for the power law case,
and when the step Eq.~(\ref{eq:step}) has been introduced in the quadratic 
potential Eq.~(\ref{eq:qp}), the small field model Eq.~(\ref{eq:sfm}) 
and the tachyon model Eq.~(\ref{eq:tm}). 
We find that the values we have obtained upon comparing the power law
case with the WMAP five and seven-year data and the QUaD and the ACBAR 
data match well with the results quoted by the WMAP~\cite{wmap-5,wmap-7} 
and the QUaD teams~\cite{quad-2009}.
Also, the results for the quadratic potential with the step agree well 
with the results quoted in the recent work~\cite{mortonson-2009}.
%%%%%%%%%%%%%%%%%%%%%%%%%%%%%%%%%%%%%%%%%%%%%%%%%%%%%%%%%%%%%%%%%%%%%%%%%%%%%%%
\begin{table}[!htb]
\begin{center}
\begin{scriptsize}
\begin{tabular}{|c|c|c|c|c|}
\hline\hline\hline
Datasets & WMAP-$5$ & WMAP-$5\,+\,$QUaD & WMAP-$5\,+\,$QUaD & WMAP-$7$\\
\cline{1-1}
Parameters & & & $+\,$ACBAR & \\ 
\hline\hline\hline
$\Omega_{\rm b}\, h^2$ & $0.0232$ & $0.0235$ & $0.0229$ & $0.0226$\\
\hline
$\Omega_{\rm c}\, h^2$ & $0.1051$ & $0.1011$ & $0.1071$ & $0.1108$\\
\hline
$\theta$ & $1.041$ & $1.043$ & $1.042$ & $1.040$\\
\hline
$\tau$ & $0.0833$ & $0.0957$ & $0.0884$ & $0.0895$\\
\hline
${\rm ln}\, \l[10^{10}\, A_{_{\rm S}}\r]$ & $3.040$ & $3.047$ 
& $3.053$ & $3.088$\\
\hline  
$n_{_{\rm S}}$ & $0.9764$ & $0.9835$ & $0.9677$ & $0.9726$\\
\hline 
$r$ & $0.3841$ & $0.4150$ & $0.0667$ & $0.1128$\\
\hline 
$n_{_{\rm T}}$ & $0.4112$ & $0.4088$ & $0.4109$ & $0.3581$\\
\hline\hline\hline
\end{tabular}
\caption{\label{tab:pl}The best fit values that we arrive at for
the input parameters upon comparing the power law primordial spectra 
Eq.~(\ref{eq:ps-pl}) with the WMAP five and seven-year, the QUaD and 
the ACBAR data sets.
We should point out that the best fit values that we have arrived at 
on using the WMAP five and seven-year data match well with the values 
quoted by the WMAP teams~\cite{wmap-5,wmap-7}.
Similarly, we find that the values we have obtained upon comparing with
the WMAP five-year and the QUaD and the ACBAR data are in good agreement 
with the results arrived at by, say, the QUaD team~\cite{quad-2009}.
Note that, while the WMAP teams~\cite{wmap-5,wmap-7} had worked with the 
pivot point of $k_{0}=0.002\; {\rm Mpc}^{-1}$, the QUaD team had set the 
pivot scale to be $k_{0}=0.05\; {\rm Mpc}^{-1}$, as we do.
However, we should clarify that, whereas the WMAP and the QUaD teams had 
imposed the consistency condition Eq.~(\ref{eq:cc}), we have not done so 
in our analysis.}
\end{scriptsize}
\end{center}
\end{table}
%%%%%%%%%%%%%%%%%%%%%%%%%%%%%%%%%%%%%%%%%%%%%%%%%%%%%%%%%%%%%%%%%%%%%%%%%%%%%%%
\begin{table}[!htb]
\begin{center}
\begin{scriptsize}
\begin{tabular}{|c|c|c|c|c|c|}
\hline\hline\hline
Models & Datasets & WMAP-$5$ & WMAP-$5$ & WMAP-$5\,+\,$QUaD 
& WMAP-$7$\\
\cline{2-2}
& Parameters & & $+\,$QUaD & $+\,$ACBAR & \\ 
\hline\hline\hline
& $\Omega_{\rm b}\, h^2$ & $0.0228$ & $0.0227$ & $0.0228$ &  $0.0224$\\
\cline{2-6}
& $\Omega_{\rm c}\, h^2$ & $0.1109$ & $0.1091$ & $0.1094$ & $0.1108$\\
\cline{2-6}
& $\theta$ & $1.041$ & $1.041$ & $1.042$ & $1.0397$\\
\cline{2-6}
Quadratic potential & $\tau$ & $0.0814$ & $0.0869$ & $0.0902$ & $0.0848$\\
\cline{2-6}
with step & ${\rm ln}\, \l[10^{10}\, m^2\r]$ & $-0.6893$ & $-0.6849$ 
& $-0.6774$ & $-0.6717$\\
\cline{2-6}
& $\alpha \times 10^4$ & $13.96$ & $15.02$ & $13.95$ & $16.06$\\
\cline{2-6}
& $\phi_0$ & $14.67$ & $14.67$ & $14.67$ & $14.67$\\
\cline{2-6}
& $\Delta\phi$ & $0.0257$ & $0.0259$ & $0.0290$ & $0.0311$\\
\hline\hline
& $\Omega_{\rm b}\, h^2$  & $0.0228$ & $0.0228$ & $0.0229$ & $0.0222$ \\
\cline{2-6}
& $\Omega_{\rm c}\, h^2$ & $0.1082$ & $0.1084$ & $0.1096$ & $0.1114$ \\
\cline{2-6}
& $\theta$ & $1.041$ & $1.041$ & $1.042$ & $1.038$ \\
\cline{2-6}
Small field model & $\tau$ & $0.0857$ & $0.0868$ & $0.0847$ & $0.0813$\\
\cline{2-6}
with step & ${\rm ln}\, \l[10^{10}\, V_{0}\r]$ & $1.684$ & $1.690$ 
& $1.689$ & $1.705$ \\
\cline{2-6}
& $-\alpha \times 10^{3}$ & $0.1153$ & $0.1371$ & $0.1701$ & $0.1569$\\
\cline{2-6}
& $\phi_{0}$ & $7.888$ & $7.887$ & $7.887$ & $7.888$\\
\cline{2-6}
& $\Delta\phi$ & $0.0070$ & $0.0076$ & $0.0089$ & $0.0090$\\
\hline\hline
& $\Omega_{\rm b}\, h^2$ & $0.0226$ &$0.0228$ &$0.0227$ & $0.0222$ \\
\cline{2-6}
& $\Omega_{\rm c}\, h^2$ &$0.1113$ &$0.1103$ &$0.1104$ & $0.1129$\\
\cline{2-6}
& $\theta$ & $1.040$ & $1.041$ & $1.042$ & $1.039$\\
\cline{2-6}
Tachyon model & $\tau$ & $0.0908$ & $0.0937$ & $0.0926$ & $0.0829$\\  
\cline{2-6} 
with step & ${\rm ln}\, \l[10^{10}\, \phi_{\ast}\r]$ & $34.51$ & $34.51$ 
& $34.51$ & $34.51$\\
\cline{2-6}
& $-\alpha$ & $0.0014$ & $0.0016$ & $0.0014$ & $0.0015$\\
\cline{2-6}
& $\phi_{0} \times 10^{-6}$ & $0.7818$ & $0.7828$ & $0.7826$ & $0.7813$\\
\cline{2-6}
& $\Delta \phi$ & $378.8$ & $371.9$ & $341.1$ & $352.4$\\
\hline\hline\hline
\end{tabular}
\caption{\label{tab:am-ws}The best fit values for the various input 
parameters corresponding to the three inflationary models with the
step.
We should point out that the best fit values for the parameters that we 
have arrived at for the quadratic potential with the step and the WMAP 
five-year data are in good agreement with the results quoted in the 
recent work~\cite{mortonson-2009}.}
\end{scriptsize}
\end{center}
\end{table}
%%%%%%%%%%%%%%%%%%%%%%%%%%%%%%%%%%%%%%%%%%%%%%%%%%%%%%%%%%%%%%%%%%%%%%%%%%%%%%%

\begin{figure}[!htb]
\begin{center}
\resizebox{360pt}{300pt}{\includegraphics{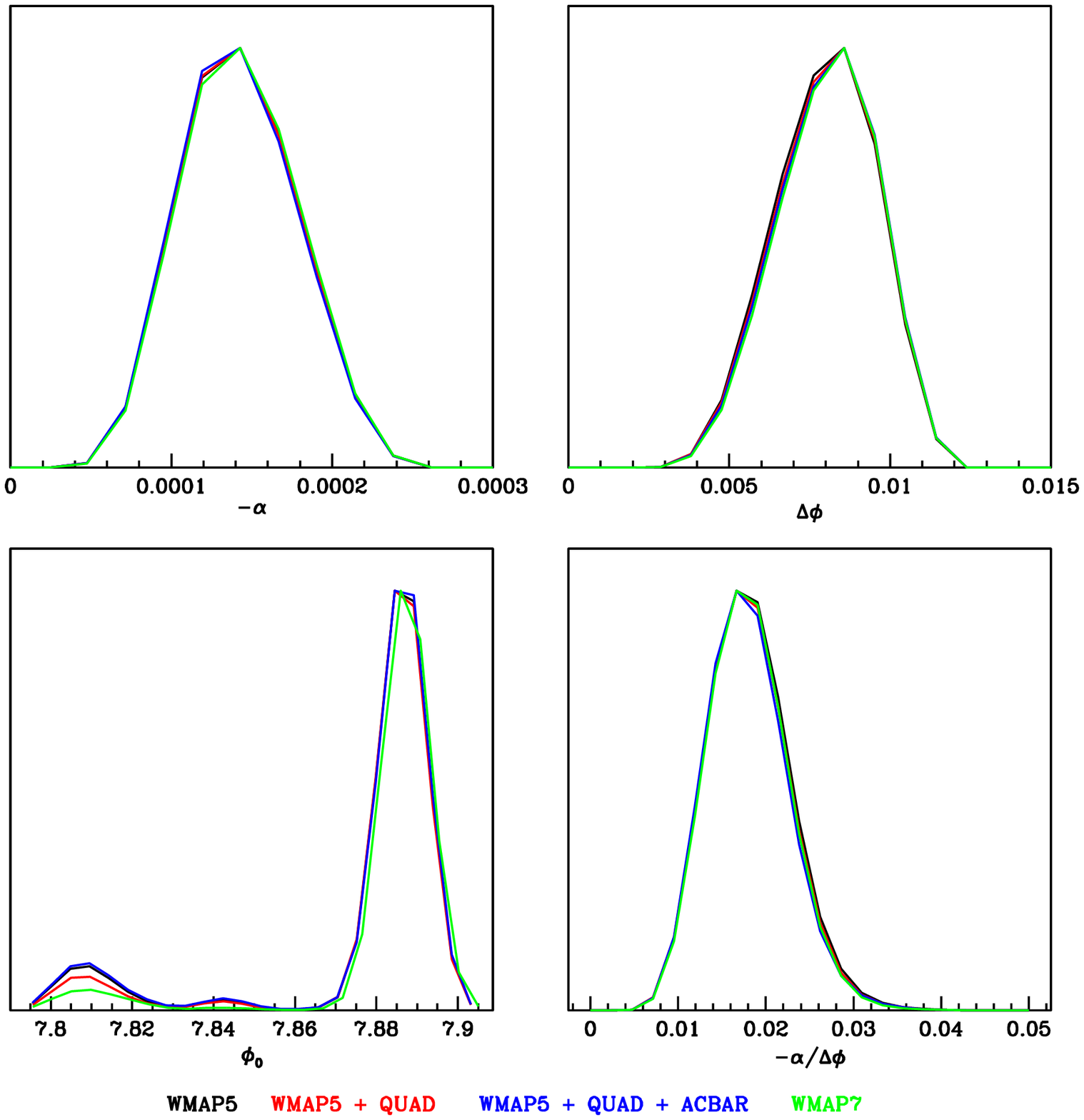}}
\end{center}
\caption{\label{fig:smf-step}
The one-dimensional likelihood distributions for the parameters describing 
the step in the case of the small field model. 
Note that, while the datasets strongly constrain the location of the 
step, the bounds are not equally tight on the height and the width of 
the step.
However, the data seems to constain the ratio of the height to the width 
of the step more tightly.}
\end{figure}

In Figure~\ref{fig:smf-step}, we have plotted the likelihood curves for 
the parameters that characterize the step in the case of the small field 
model.
In the figure, we have also included the one-dimensional likelihood for 
the ratio $\l(\alpha/\Delta\phi\r)$ of the step.
(Note that, in the figure, WMAP-$5$ and WMAP-$7$ refer to the WMAP five and
seven-year data, respectively.)
We find that, while the data constrain the location of the step quite well 
in all the models, the bounds on the height and the width of the step are 
not equally tight. 
Interestingly, we find that the data constrains the ratio of the height to
the width of the step fairly tightly.
%%%%%%%%%%%%%%%%%%%%%%%%%%%%%%%%%%%%%%%%%%%%%%%%%%%%%%%%%%%%%%%%%%%%%%%%%%%%%%%
%%%%%%%%%%%%%%%%%%%%%%%%%%%%%%%%%%%%%%%%%%%%%%%%%%%%%%%%%%%%%%%%%%%%%%%%%%%%%%%

\subsection{The effective least squares parameter $\chi_{\rm eff}^2$}

In Table~\ref{tab:chsq}, we have listed the least squares parameter 
$\chi_{\rm eff}^{2}$ for all the different models and datasets of our 
interest.
%%%%%%%%%%%%%%%%%%%%%%%%%%%%%%%%%%%%%%%%%%%%%%%%%%%%%%%%%%%%%%%%%%%%%%%%%$$$$$$
\begin{table}[!htb]
\begin{center}
\begin{scriptsize}
\begin{tabular}{|c|c|c|c|c|}
\hline\hline\hline
Datasets & WMAP-$5$ & WMAP-$5$ & WMAP-$5\,+\,$QUaD & WMAP-$7$\\
\cline{1-1}
Models & & $\,+\,$QUaD & $\,+\,$ACBAR &\\
\hline\hline
Power law case $(4,4)$ & $2658.40$ & $2757.34$ & $2779.12$ 
& $7474.48$\\
\hline
Quadratic potential $(1,1)$ & $2658.22$ ($-0.18$)& $2757.54$ ($+0.20$)
& $2779.02$ ($-0.10$) & $7474.78$ ($+0.30$)\\
\hline
Quadratic potential$\,+\,$step $(4,4)$ & $2651.00$ ($-7.40$) 
& $2750.38$ ($-6.96$) & $ 2771.72$ ($-7.40$) & $7466.28$ ($-8.20$)\\
\hline
Small field model $(3,1)$ & $2658.26$ ($-0.14$) & $2757.46$ ($+0.12$)
& $2779.06$ ($-0.06$) & $7474.78$ ($+0.30$)\\
\hline
Small field model$\,+\,$step $(6,4)$ & $2650.96$ ($-7.44$) 
& $2750.26$ ($-7.08$) & $2771.92$ ($-7.20$) & $7466.00$ ($-8.48$)\\
\hline
Tachyonic model $(2,1)$ & $2658.26$ ($-0.14$) & $2757.60$ ($+0.26$) 
& $2779.10$ ($-0.02$) & $7474.56$ ($+0.08$) \\
\hline
Tachyonic model$\,+\,$step $(5,4)$ & $2651.14$ ($-7.26$) 
& $2750.50$ ($-6.84$) & $2772.06$ ($-7.06$) & $7465.92$ ($-8.56$)\\
\hline\hline\hline
\end{tabular}
\caption{\label{tab:chsq}The $\chi_{\rm eff}^{2}$ for the different 
models and datasets that we have considered. 
The two quantities that appear within the brackets in the leftmost column 
indicate the number of inflationary parameters available in the different 
models and the number of parameters that we have varied when comparing the
models against the data, in that order.
The quantities within the brackets in the remaining columns indicate the
difference in the $\chi_{\rm eff}^{2}$ between the model and the power
law case for that dataset, with a negative value indicating an improvement
in the fit.
Note that, as we had mentioned earlier, the Gibbs approach in the WMAP
likelihood code has been used to calculate the $\chi_{\rm eff}^{2}$ for 
the CMB $TT$ spectrum at the low multipoles (i.e. for $\ell 
<32$)~\cite{wmap-5,wmap-7}.
Without the step, all the inflationary models perform just as well as the
power law case.
And, evidently, the introduction of the step reduces the $\chi_{\rm eff}^{2}$ 
by about $7$-$9$ in {\it all}\/ the cases.}
\end{scriptsize}
\end{center}
\end{table}
%%%%%%%%%%%%%%%%%%%%%%%%%%%%%%%%%%%%%%%%%%%%%%%%%%%%%%%%%%%%%%%%%%%%%%%%%%%%%%%
It is clear from the table that the presence of the step leads to a 
reduction in $\chi_{\rm eff}^{2}$ by about $7$-$9$ in all the three 
inflationary models that we have considered.
Also, note that such an improvement is achieved in all the datasets 
that we have compared the models with.

\begin{figure}[!htb]
\begin{center}
\hskip 15pt
\resizebox{210pt}{145pt}{\includegraphics{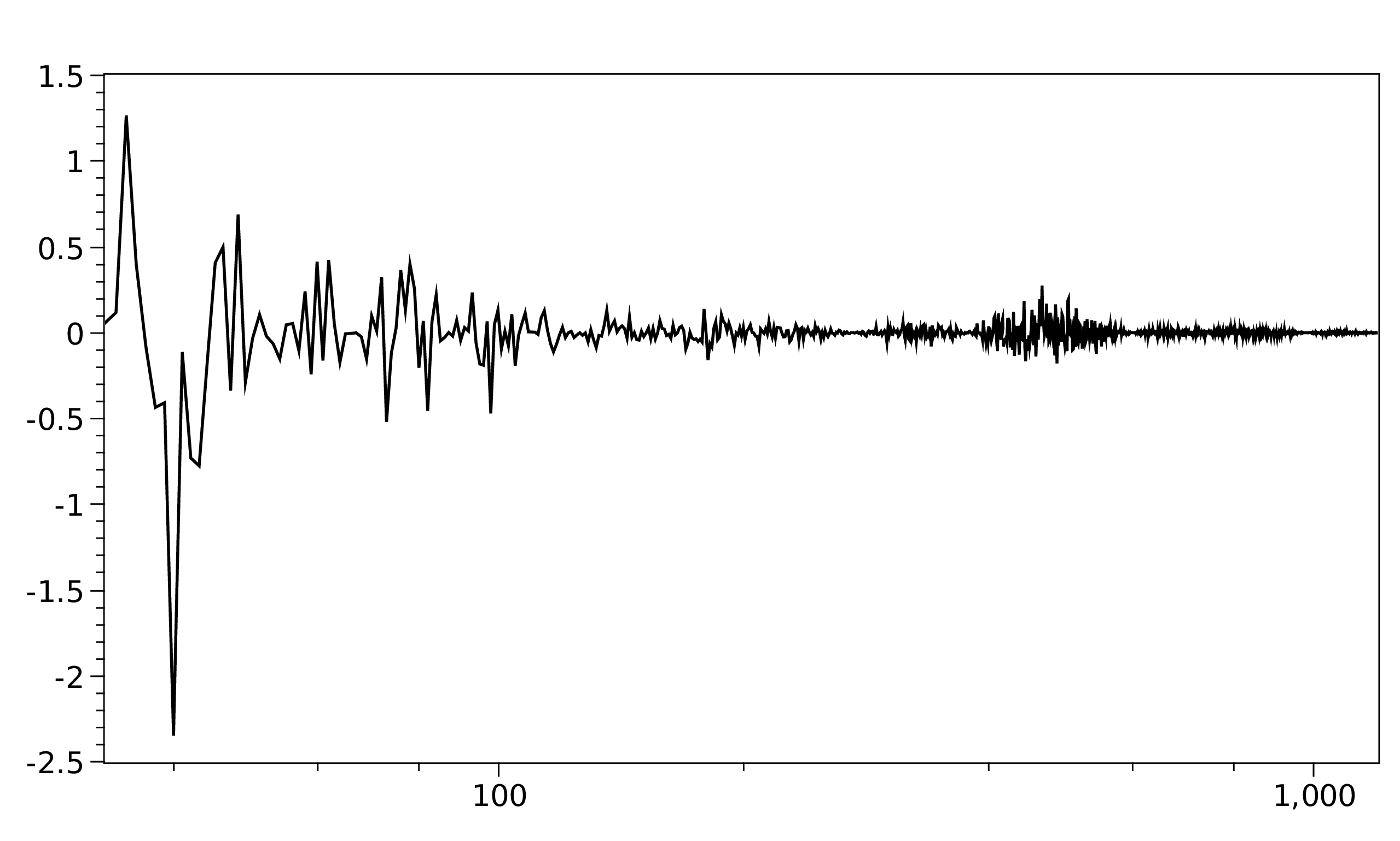}}
\hskip 3pt
\resizebox{210pt}{145pt}{\includegraphics{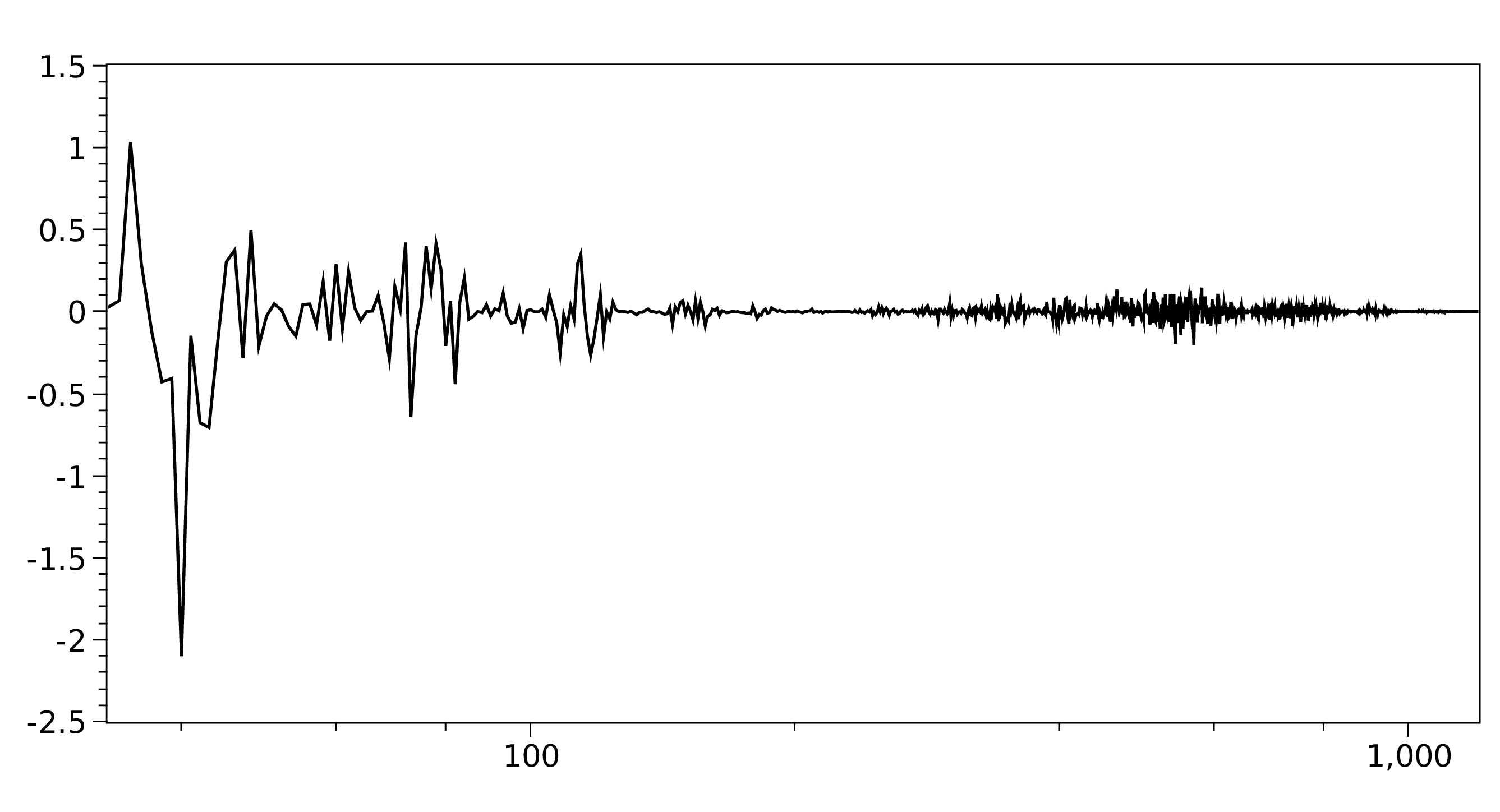}}
\vskip -102 true pt 
\hskip -428 true pt \rotatebox{90}{$\Delta\chi_{\rm eff}^{2}$}
\vskip 65 pt
\hskip 50 true pt $\ell$ \hskip 210 true pt $\ell$
\end{center}
\caption{\label{fig:chsqdiff}The difference in $\chi_{\rm eff}^{2}$ for
the WMAP seven-year data with and without the step has been plotted as a 
function of the multipole moment for $\ell>32$.
The plot on the left corresponds to the quadratic potential, while the one 
on the right is for the small field model.
The two figures are strikingly similar, and it is clear that the improvement 
in $\chi_{\rm eff}^{2}$ occurs near $\ell\simeq 40$ in both the cases.
We find that the corresponding result for the tachyon model behaves in
essentially the same fashion.}
\end{figure}

When we compare the contribution to the $\chi_{\rm eff}^{2}$ at the low 
multipoles (i.e. up to $\ell=32$, see Refs.~\cite{wmap-5,wmap-7}) from the 
output of the WMAP likelihood code for, say, the WMAP seven-year data, we 
find that the introduction of the step in the inflaton potential reduces the 
$\chi_{\rm eff}^{2}$ for the $TT$ data over this range by about $5$-$6$ in 
all the cases.
In Figure~\ref{fig:chsqdiff}, we have plotted the difference in 
$\chi_{\rm eff}^{2}$ with and without the step, as a function of 
the multipoles when $\ell>32$, for the WMAP seven-year data. 
We have plotted the difference in $\chi_{\rm eff}^{2}$ for the cases of 
the quadratic potential and the small field model.
%%%%%%%%%%%%%%%%%%%%%%%%%%%%%%%%%%%%%%%%%%%%%%%%%%%%%%%%%%%%%%%%%%%%%%%%%%%%%%%
%%%%%%%%%%%%%%%%%%%%%%%%%%%%%%%%%%%%%%%%%%%%%%%%%%%%%%%%%%%%%%%%%%%%%%%%%%%%%%%
It is clear from the figure that the additional improvement by about $2$
arises due to a better fit near $\ell=40$. 
(There also seems to be a `loss' of about unity in $\chi_{\rm eff}^{2}$ at 
$\ell\lesssim 40$.)
In other words, the step essentially improves the fit to the data at the 
lower multipoles.
This point will be further evident in the following section, wherein we 
discuss the resulting CMB angular power spectrum.

%%%%%%%%%%%%%%%%%%%%%%%%%%%%%%%%%%%%%%%%%%%%%%%%%%%%%%%%%%%%%%%%%%%%%%%%%%%%%%%

\section{The scalar and the CMB angular power spectra}\label{sec:sps}

\begin{figure}[!htb]
\begin{center}
\resizebox{160pt}{100pt}{\includegraphics{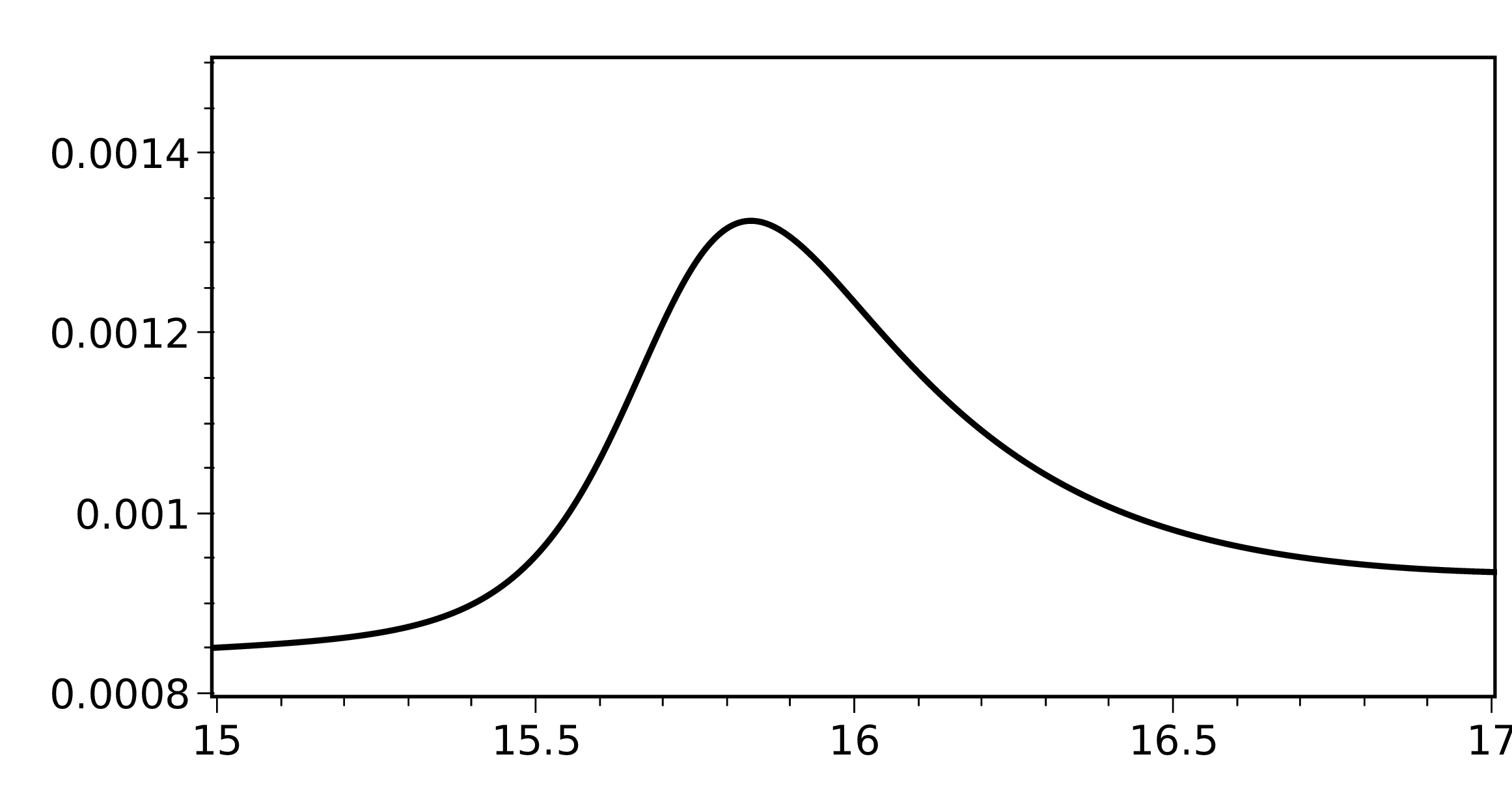}}
\hskip 40pt
\resizebox{160pt}{100pt}{\includegraphics{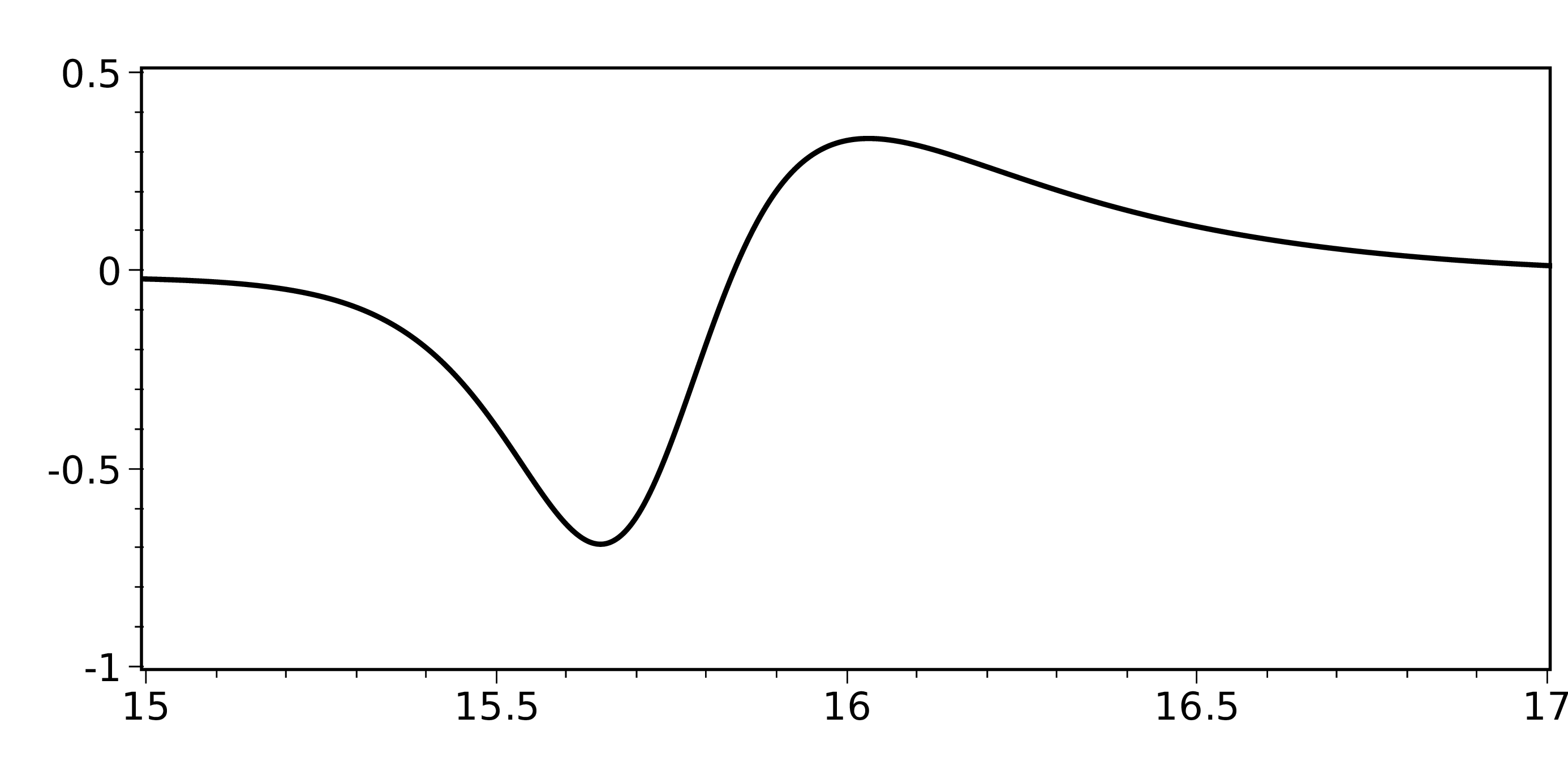}}
\vskip -80 true pt 
\hskip -170 true pt $\epsilon$
\hskip 200 true pt $\eta$
\vskip 60 pt
\hskip 15 true pt $N$ \hskip 185 true pt $N$  
\caption{\label{fig:dsri}Typical evolution of the first two slow roll 
parameters $\epsilon$ and $\eta$ with the introduction of the step for 
the three inflationary models that we have considered.
We have plotted above the evolution of the slow roll parameters as a 
function of the e-folds $N$ for the small field model around the time 
when the field crosses the step in the potential.}
\end{center}
\end{figure}

\begin{figure}[!htb]
\begin{center}
\resizebox{360pt}{240pt}{\includegraphics{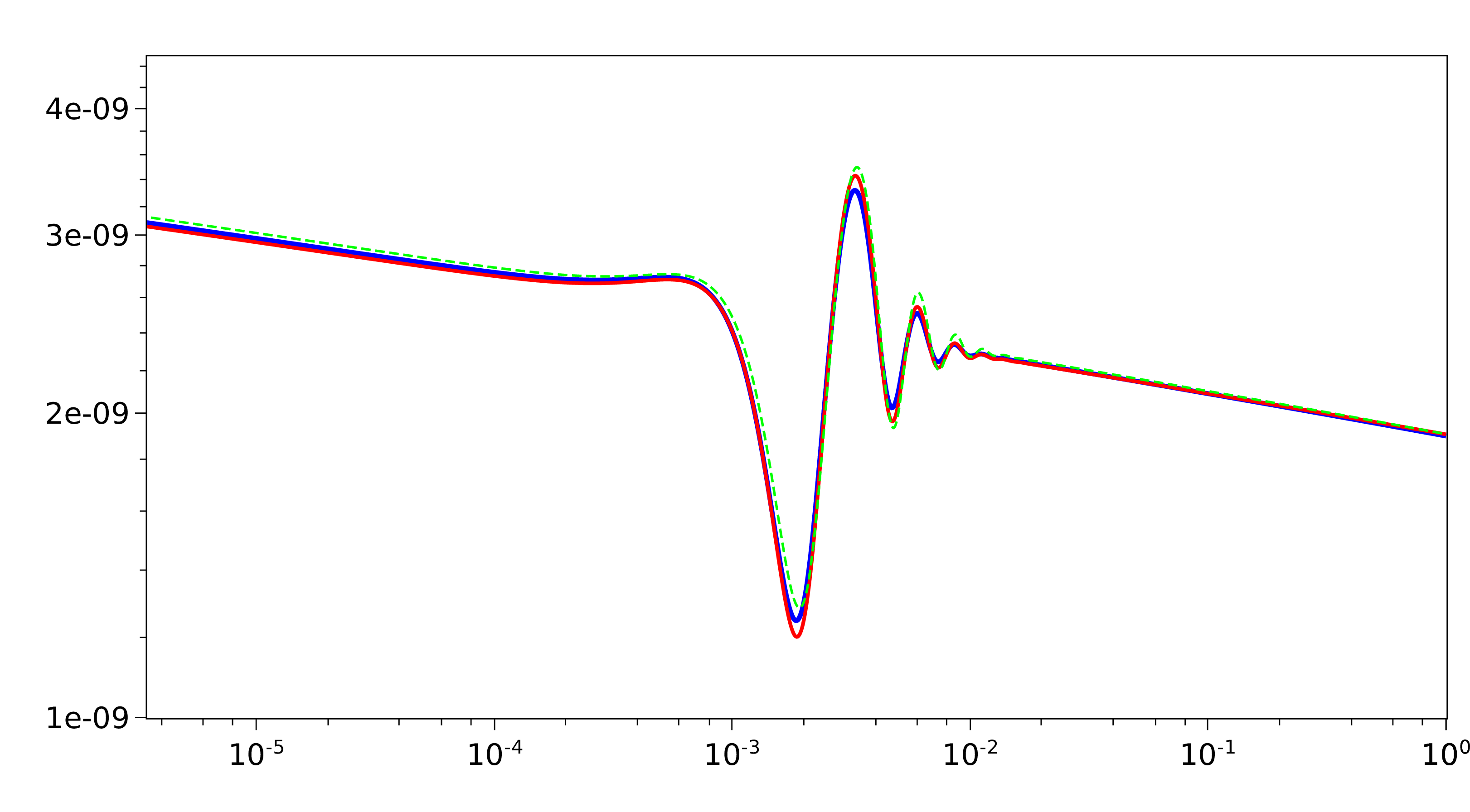}}
\vskip -158 true pt 
\hskip -368 true pt ${\cal P}_{_{\rm S}}(k)$
\vskip 128 pt
\hskip 40 true pt $k$ 
\caption{\label{fig:sps}The scalar power spectra corresponding to the 
best fit values of the WMAP seven-year data for the inflationary models 
with the step.
The solid blue, the solid red, and the dashed green curves describe the 
scalar power spectra in the cases of the quadratic potential, the small 
field model, and the tachyon model, respectively.
Evidently, the three spectra are hardly distinguishable.
And, obviously, the oscillations will not arise in the absence of the 
step.}
\end{center}
\end{figure}

As we had mentioned in the opening section, the introduction 
of the step leads to a small deviation from slow roll 
inflation~\cite{covi-2006-2007,mortonson-2009}.
We have illustrated this behavior in Figure~\ref{fig:dsri}, wherein 
we have plotted the evolution of the first two slow roll parameters 
$\epsilon$ and $\eta$ around the time when the field crosses the step 
in the small field model. 
We find that essentially the same behavior arises in all the three
inflationary models that we have considered (for the definition of 
these slow roll parameters in the case of the canonical scalar field 
and the tachyon models, see, for instance, Refs.~\cite{sriram-2009} 
and~\cite{steer-2004}, respectively).
%%%%%%%%%%%%%%%%%%%%%%%%%%%%%%%%%%%%%%%%%%%%%%%%%%%%%%%%%%%%%%%%%%%%%%%%%%%%%%%
%%%%%%%%%%%%%%%%%%%%%%%%%%%%%%%%%%%%%%%%%%%%%%%%%%%%%%%%%%%%%%%%%%%%%%%%%%%%%%%
The small deviation from slow roll inflation leads to a burst of 
oscillations superimposed on the otherwise nearly scale invariant 
scalar power spectrum, as we have illustrated in Figure~\ref{fig:sps}.
%%%%%%%%%%%%%%%%%%%%%%%%%%%%%%%%%%%%%%%%%%%%%%%%%%%%%%%%%%%%%%%%%%%%%%%%%%%%%%%
%%%%%%%%%%%%%%%%%%%%%%%%%%%%%%%%%%%%%%%%%%%%%%%%%%%%%%%%%%%%%%%%%%%%%%%%%%%%%%% 
We should add that, since the deviation from slow roll is relatively
small, the introduction of the step hardly affects the tensor spectrum.
It remains nearly scale invariant in all the cases\footnote{We should 
mention that, as the effects of the step in the potential on the tensor 
spectrum are rather small, we could have as well worked with a suitable 
scale invariant amplitude for including the tensors~\cite{mortonson-2009}.}.
At the point $k=0.05\; {\rm Mpc}^{-1}$, we find the tensor-to-scalar 
ratio $r$ to be about $0.16$, $0.016$ and $0.16$ in the cases of the 
quadratic potential, the small field and the tachyon models, 
respectively.

\begin{figure}[!htb]
\begin{center}
\hskip 25pt
\resizebox{400pt}{300pt}{\includegraphics{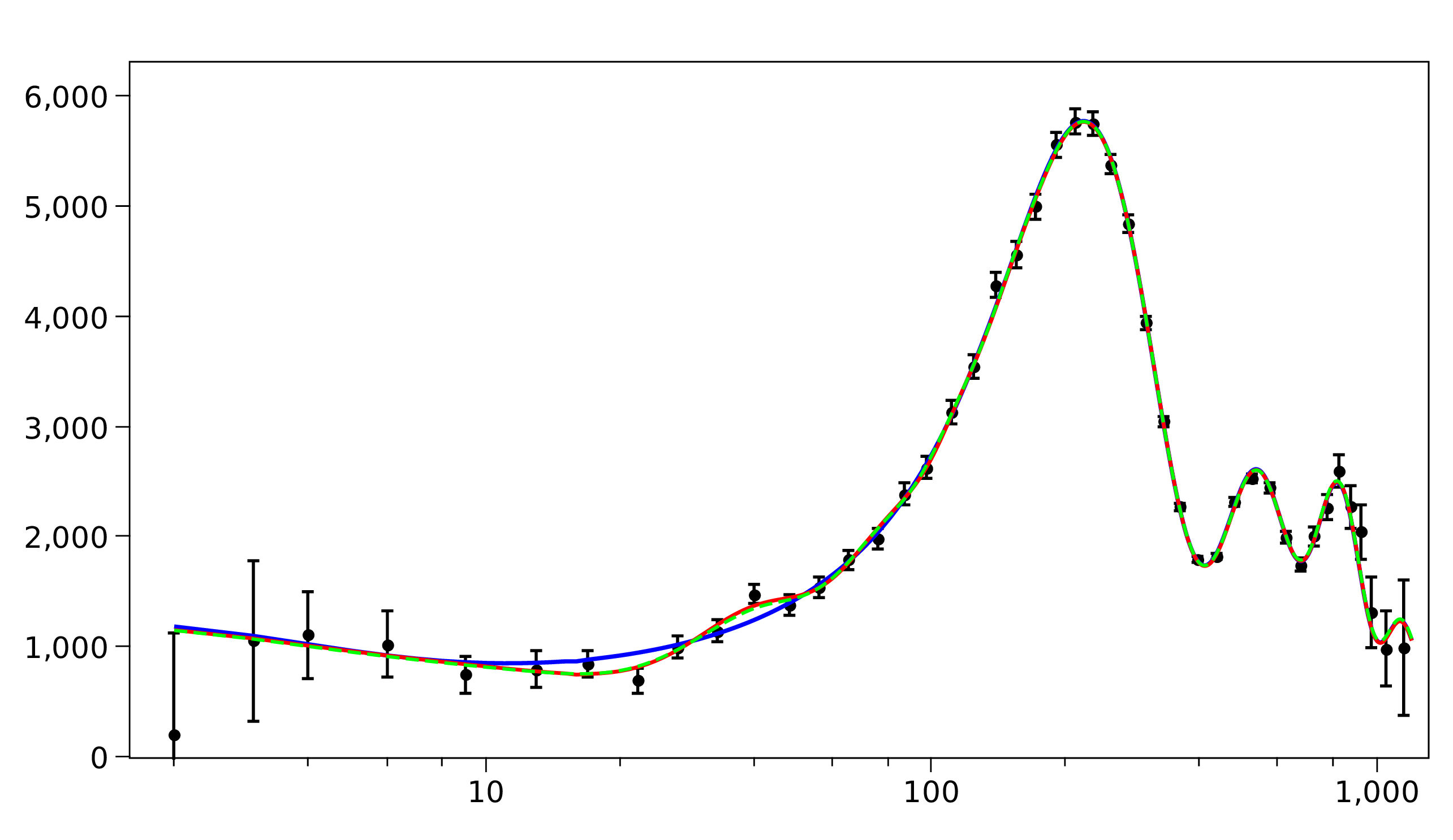}}
\vskip -230 true pt \hskip -400 true pt 
\rotatebox{90}{$\l[\ell\, (\ell+1)\;
C_{\ell}^{TT}/(2\,\pi)\r]\; \mu\, {\rm K}^{2}$}
\vskip 85 true pt \hskip 40 true pt $\ell$
\caption{\label{fig:cl1}The CMB $TT$ angular power spectra corresponding 
to the best fit values of the inflationary models for the WMAP seven-year 
data without and with the step.
The solid blue and the solid red curves correspond to the quadratic potential
without and with the step, respectively.
The dashed green curve corresponds to the best fit small field model with the
step.
We find that the results for the tachyon model behave in exactly the same 
fashion.
The black dots with error bars denote the WMAP seven-year data.
It is visually evident that, with the introduction of the step, the models 
lead to a better fit to the data near the multipole moments of $\ell=22$ and
$40$.}
\end{center}
\end{figure}

\begin{figure}[!htb]
\begin{center}
\hskip 25pt
\resizebox{404pt}{292pt}{\includegraphics{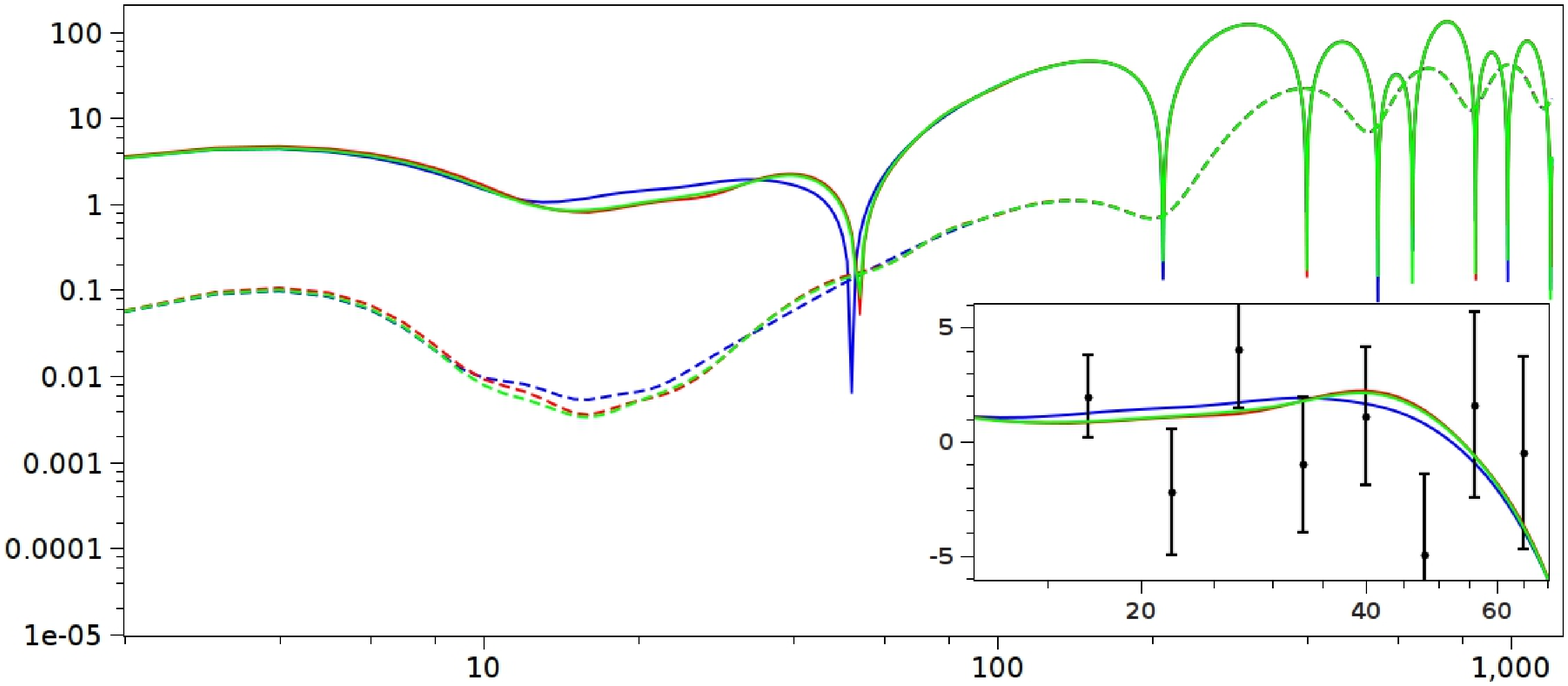}}
\vskip -230 true pt \hskip -400 true pt 
\rotatebox{90}{$\l[\ell\, (\ell+1)\;
C_{\ell}/(2\,\pi)\r]\; \mu\, {\rm K}^{2}$}
\vskip 95 true pt \hskip 40 true pt $\ell$
\caption{\label{fig:cl2}The CMB $TE$ and the $EE$ angular power spectra 
corresponding to the best fit values of the inflationary models for the 
WMAP seven-year data without and with the step.
The solid blue, red and green curves represent the $TE$ spectrum (actually, 
its magnitude) for the quadratic potential without and with the step, and 
the small field model with the step, respectively.
The dashed curves denote the corresponding $EE$ angular power spectra.
The inset indicates the behavior of the $TE$ angular power spectra 
against the WMAP seven-year data over the multipole 
interval where the discrimination is maximal.}
\end{center}
\end{figure}

The burst of oscillations in the scalar power spectrum in turn results 
in a feature in the CMB $TT$ angular power spectrum, which leads to the 
improvement in the fit to the data at the lower multipoles. 
This behavior is evident in Figure~\ref{fig:cl1} wherein we have plotted 
the CMB $TT$ angular power spectra for the quadratic potential without 
and with the step, and for the small field model with the step.
%%%%%%%%%%%%%%%%%%%%%%%%%%%%%%%%%%%%%%%%%%%%%%%%%%%%%%%%%%%%%%%%%%%%%%%%%%%%%%%
%%%%%%%%%%%%%%%%%%%%%%%%%%%%%%%%%%%%%%%%%%%%%%%%%%%%%%%%%%%%%%%%%%%%%%%%%%%%%%%
In Figure~\ref{fig:cl2}, we have plotted the corresponding $TE$ and 
$EE$ angular power spectra.
%%%%%%%%%%%%%%%%%%%%%%%%%%%%%%%%%%%%%%%%%%%%%%%%%%%%%%%%%%%%%%%%%%%%%%%%%%%%%%%

%%%%%%%%%%%%%%%%%%%%%%%%%%%%%%%%%%%%%%%%%%%%%%%%%%%%%%%%%%%%%%%%%%%%%%%%%%%%%%%

\section{Summary and discussion}\label{sec:summary}

In this work, we have investigated the effects of introducing a step 
in certain inflationary models.
In addition to revisiting the case of the quadratic potential that has
been considered earlier, we have studied the effects of the step in a
small field model and a tachyon model. 
The introduction of the step leads to a small deviation from slow roll
inflation, which results in a burst of oscillations in the scalar power
spectrum.
These oscillations, in turn, leave their imprints as specific features 
in the CMB angular power spectrum.
Actually, we have also evaluated the tensor power spectrum exactly, and 
have included it in our analysis.
We believe that this is a timely effort considering the fact that 
results from, say, the ongoing PLANCK mission might necessitate 
such an analysis.
Upon comparing the inflationary models with the WMAP, the QUaD and the 
ACBAR data, we find that, with the step, all the models lead to an 
improvement in $\chi_{\rm eff}^{2}$ by about $7$-$9$ over the smooth, 
nearly scale invariant, slow roll spectrum, at the expense of three 
additional parameters describing the location, the height and the width 
of the step in the inflaton potential.
The output of the WMAP likelihood code and a plot of the difference in 
$\chi_{\rm eff}^{2}$ with and without the step clearly illustrate that 
the improvement occurs because of a better fit to the data at the lower 
multipoles due to the presence of the step.
Evidently, if future observations indicate that the amplitude of the 
tensors are rather small, then the quadratic potential and the 
tachyon model will be ruled out, while a suitable small field model 
with a step can be expected to perform well against the data.

The introduction of the step in an inflationary model can possibly be 
viewed as an abrupt change in a potential parameter~\cite{adams-2001}.
But, it has to be admitted that it is rather ad-hoc, and one needs to
explore the generation of features and a resulting improvement in the 
fit in better motivated inflationary models.
Two field models offer such a possibility.
For instance, with suitably chosen parameters, the two field models can 
easily lead to a brief departure from slow roll inflation (in this context, 
see, for example, Refs.~\cite{joy-2008-2009,cline-2003,hunt-2004-2007}). 
However, iso-curvature perturbations arise whenever more than one 
field is involved~\cite{gordon-2000,tsujikawa-2003-2005}, and they
need to be carefully taken into account when comparing these models 
with the data.
It will be a worthwhile effort to systematically explore the two field 
models, including the effects due to the iso-curvature perturbations, in 
an attempt to fit the outliers near the multipole moments of $\ell=22$ 
and $40$.
 
Over the last few years, it has been recognized that primordial
non-Gaussianity can act as a powerful observational tool that can 
help us discriminate further between the various inflationary models. 
For example, it has been shown that slow roll inflation driven by the 
canonical scalar fields leads only to a small amount of 
non-Gaussianity~\cite{maldacena-2003}.
However, recent analysis of the CMB data seem to suggest 
that the extent of primordial non-Gaussianity may possibly 
be large (see, for instance, Refs.~\cite{wmap-7,smith-2009}).  
It is known that models which lead to features, such as the ones 
we have considered here, also generate a reasonably large 
non-Gaussianity (see, for example, Refs.~\cite{chen-2007-2008}).
While the different models that we have considered in this work lead 
to virtually the same scalar power spectrum and almost the same extent 
of improvement in the fit (i.e. with the introduction of the step) to
the CMB data, it is important to examine whether they lead to the same 
extent of non-Gaussianity as well.
We are currently investigating such issues.

%%%%%%%%%%%%%%%%%%%%%%%%%%%%%%%%%%%%%%%%%%%%%%%%%%%%%%%%%%%%%%%%%%%%%%%%%%%%%%%%

\section*{Acknowledgments}

We wish to thank Hiranya Peiris for extensive exchanges over e-mail 
on a few different issues related to the comparison of models with 
the data.
RKJ and LS would like to thank Pravabati Chingangbam for discussions 
on related topics.
DKH and LS would like to thank Jerome Martin for discussions.
DKH also wishes to thank Sanjoy Biswas, Joydeep Chakrabortty, and 
Tirthankar Roy Choudhury for various help on numerical matters.
MA wishes to thank Antony Lewis for valuable suggestions.
MA would also like to thank the Harish-Chandra Research Institute, 
Allahabad, India, for hospitality, where part of this work was carried 
out.
We would also like to acknowledge the use of the high performance
computing facilities at the Harish-Chandra Research Institute,
Allahabad, India, as well as at the Inter-University Centre for 
Astronomy and Astrophysics, Pune, India.
RKJ acknowledges financial support from the Swiss National Science  
Foundation.
TS and MA acknowledge support from the Swarnajayanti Fellowship, 
Department of Science and Technology, India.
Finally, we acknowledge the use of the CosmoMC package~\cite{cosmomc},
and the data products provided by the WMAP science team~\cite{lambda},
the QUaD and the ACBAR missions.

%%%%%%%%%%%%%%%%%%%%%%%%%%%%%%%%%%%%%%%%%%%%%%%%%%%%%%%%%%%%%%%%%%%%%%%%%%%%%%%

%%%%%%%%%%%%%%%%%%%%%%%%%%%%%%%%%%%%%%%%%%%%%%%%%%%%%%%%%%%%%%%%%%%%%%%%%%%%%%%
\end{document}